\documentclass{article}
\usepackage{amsmath}
\usepackage{graphicx}
\usepackage{subfigure}
\usepackage{hyperref}
\usepackage{natbib}
\usepackage{url} 
\usepackage{amssymb}
\usepackage{xcolor}
\usepackage{mathabx}
\usepackage{comment}
\usepackage{tabularx}
\usepackage{booktabs}
\usepackage{enumitem}
\usepackage[ruled]{algorithm2e}
\usepackage{multirow}%
\usepackage{amsmath,amssymb,amsfonts}%
\usepackage{amsthm}%
\usepackage{mathrsfs}%
\usepackage[title]{appendix}%
\usepackage{textcomp}%
\usepackage{manyfoot}%
\usepackage{booktabs}%
\usepackage{algorithmicx}%
\usepackage{algpseudocode}%
\usepackage{listings}%
\usepackage{subfigure}

\usepackage{tikz}

\usetikzlibrary{shapes,decorations,arrows,calc,arrows.meta,fit,positioning}

\usepackage{pgfplots}

\usepackage{xcolor}
\tikzset{
	-Latex,auto,node distance =1 cm and 1 cm,semithick,
	state/.style ={circle, draw, minimum width = 0.7 cm},
	box/.style ={rectangle, draw, minimum width = 0.7 cm, fill=lightgray},
	point/.style = {circle, draw, inner sep=0.08cm,fill,node contents={}},
	bidirected/.style={Latex-Latex,dashed},
	el/.style = {inner sep=3pt, align=left, sloped}
	
}

\graphicspath{{Figs/}}

\makeatother

\begin{document}

\def\spacingset#1{\renewcommand{\baselinestretch}%
{#1}\small\normalsize} \spacingset{1}


  \title{\bf An Infinite BART Model}
  \author{Marco Battiston\thanks{School of Mathematical Sciences, Lancaster University, UK}, \hspace{.2cm}
Yu Luo\thanks{Department of Mathematics, King's College London, U.K.} \\
  }
 \date{ }
  \maketitle

\bigskip
\begin{abstract}
Bayesian additive regression trees (BART) is a popular Bayesian ensemble model used in regression and classification analysis. Under this modeling framework, the regression function is approximated by an ensemble of decision trees, interpreted as weak learners. In this work, we propose a generalization of the BART model that has two main features: first, it automatically selects the number of decision trees using the given data; second, the model allows clusters of observations to have different regression functions since each data point can only use a selection of weak learners, instead of all of them. This model generalization is accomplished by including a binary weight matrix in the conditional distribution of the response variable, which activates only a specific subset of decision trees for each observation. Such a matrix is endowed with an Indian Buffet process prior, and sampled within the MCMC sampler, together with the other BART parameters. We then compare the Infinite BART model with the classic one on simulated and real datasets. Specifically,  we provide examples illustrating variable importance, partial dependence and causal estimation.
\end{abstract}

\noindent%
{\it Keywords:}   Bayesian Additive Regression Trees, Non-parametric Bayesian Regression, Indian Buffet Process, Variable Importance, Casual Estimation.
\vfill

\newpage
\spacingset{1.5}

\section{Introduction} \label{sec:introduction}

Bayesian Additive Regression Trees (BART) is a popular Bayesian ensemble method used in regression and classification analysis. It was  initially introduced in \citet{CMG(10)} as the generalization of Bayesian classification and regression tree (CART) model \citep{CGM(98),DMS(98)}. The CART model fits a single, usually deep, decision tree to the data, which is endowed with a prior distribution and then learned from the data through Markov chain Monte Carlo (MCMC). On the other hand, BART uses a sum-of-trees to approximate the regression function, averaging  over a set of many shallow decision trees. These can be interpreted as weak learners and capture different features of the data. Therefore, BART provides a flexible Bayesian approach to regression analysis, with often strong predictive capabilities, without any stringent parametric assumptions. 

There are numerous variants of  the classic BART model in the literature, each designed to handle different data types and tailored to a wide range of applications. For example, in \citet{CMG(10)}, an extension of the classic regression model was proposed to accommodate binary outcomes, using the probit augmentation of \citet{AC(93)}. \citet{M(21)} explains how BART can be adapted for categorical and count outcomes, including extensions based on log-linear models, multinomial logistic formulations, and count regression approaches that account for zero-inflation and over-dispersion. Heteroskedastic BART is examined in \citet{GLLMS(19)}. Extensions of the classic BART framework that incorporate priors on split rules designed for high-dimensional settings have been proposed in \citet{L(18)} and \citet{LY(18)}. Applications of BART to survival analysis are presented in \citet{BBBSAD(10)} and \citet{SLML(16)}, while its use in causal inference is discussed in \citet{H(11)} and \citet{HS(13)}. For a comprehensive overview of BART and its broad range of applications, see \citet{HLM(20)}.

In this work, we propose an extension of BART, termed as \textit{infinite BART}, which is characterized by two main characteristics. Firstly, it does not require practitioners to pre-specify the number of trees in the model; instead, the appropriate number is automatically determined based on the observed data. This is achieved by permitting a potentially infinite number of decision trees in the regression function, along with a binary weight matrix that activates only a subset of trees for each data point, determined by the observed data. The unknown weight matrix is assigned with an Indian Buffet Process (IBP), a nonparametric prior for binary matrices that is widely used in machine learning and was originally proposed by \citet{GG(05)} for factor analysis.

{The second important characteristic is that infinite BART enables (soft) clustering among data points, allowing different clusters to have heterogeneous regression functions. In applications, this situation occurs naturally when some significant factor variables are not included among the set of explanatory variables. As a toy example, assume that the `true' relation was $y=\beta_{0}+\beta_{1}x_{1}+\beta_{2}x_{2}+\beta_{4}x_{1}x_{2}+\epsilon$, with $x_{1}$ continuous and $x_{2}$ a dummy variable for a binary factor. However, assume also that information on $x_{2}$ was not available in the dataset. If we performed the regression of $y$ only on $x_{1}$, the model would need to approximate two heterogeneous regression functions, $y=\beta_{0}+\beta_{1}x_{1}+\epsilon$ for the group of observations with $x_{2}=0$ and $y=(\beta_{0}+\beta_{2})+(\beta_{1}+\beta_{4})x_{1}+\epsilon$ for the other group of observations with $x_{2}=1$, without knowing to which group ($x_{i,2}=0$ or $x_{i,2}=1$) each $y_{i}$ belongs to. With more than one unobserved factor variable, soft clustered regression functions can occur at different combinations of factor levels. Given that real-world datasets may not contain all variables that influence the dependent variable, soft clustering regression can arise naturally in many applications. Within infinite BART, the soft clustering of observations is modeled through the associated latent weight matrix.}

The proposed infinite BART model employs the three-parameter version of the IBP, introduced in \citet{TG(09)}, as the prior for the weight matrix. Given this prior choice, we introduce a MCMC sampler to perform posterior inference for model parameters. The proposed model can address common regression statistical tasks in a similar fashion as in classic BART framework. Specifically, we will discuss: point estimation and credible intervals for the regression function $f(x)$, prediction of out-of-sample observations, variable selection and variable importance, estimation of partial dependence and average casual effects. For clarity, we focus here on extending classic BART for regression with Gaussian noise; however, similar extensions can be applied to other BART variants, such as BART for binary outcomes.

The remainder of the paper is organized as follows. Section~\ref{sec:background} provides a brief background, with the classic BART model in subsection~\ref{sec:BART} and the IBP in subsection~\ref{sec:IBP}. It also establishes the notation used throughout the rest of the paper. Section~\ref{sec:InfiniteBART} describes the Infinite BART model. Specifically, subsection~\ref{sec:model} introduces the model and its priors, subsection~\ref{sec:mcmc} describes the MCMC sampler for posterior inference, and subsection~\ref{sec:tasks} explains how common statistical tasks can be addressed using infinite BART. Section~\ref{sec:applications} presents some examples of applications of infinite BART in both simulated and real datasets. Finally,  Section~\ref{sec:discussion} presents some concluding remarks and future research directions.

\section{Background} \label{sec:background}

\subsection{BART model} \label{sec:BART}

BART is a popular Bayesian model for regression and classification analysis, initially introduced in \citet{CMG(10)}. It models the regression function using an ensemble of decision regression trees and extends the Bayesian CART model \citep{CGM(98)}. In this section, we provide a brief review of the classic BART model and establish the notation used throughout. In particular, we focus on the continuous response variable with Gaussian errors setting. 

Let $\mathbf{Y}=(Y_{1},\ldots,Y_{n})$ denote the $\mathbb{R}^{n}$-dimensional response vector, including the dependent variable of $n$ observations. Each response variable, $Y_{i}$, is associated to a vector of explanatory variables $\textbf{X}_{i}=(X_{i,1},\ldots,X_{i,p})$, which may include both continuous and categorical variables. 
In the BART model, the conditional distribution of the response observation $Y_{i}$, given its explanatory variables $\textbf{X}_{i}$, is modeled through a non-parametric regression model, 
\begin{equation} \label{eq:regression} Y_{i}=f(\textbf{X}_{i})+\epsilon_{i}
\end{equation}
with $\epsilon_{i} \stackrel{\text{iid}}{\sim}N(0,\sigma^2)$. The unknown regression function $f:\mathbb{R}^{p}\to \mathbb{R}$ in \eqref{eq:regression} is assumed to be, or at least well-approximated by, an ensemble of $K$ decision trees,
\begin{equation} \label{eq:BART}
	f(\textbf{X}_{i})=\sum_{k=1}^{K}g(\textbf{X}_{i};\mathcal{T}_{k},\pmb{\mu}_{k})
\end{equation}
where $g(\textbf{X}_{i};\mathcal{T}_{k},\pmb{\mu}_{k})$ is the regression function of a decision tree with tree $\mathcal{T}_{k}$ and leaves parameters $\pmb{\mu}_{k}$. More specifically, 
\begin{enumerate}
	\item Let $\mathcal{T}_{k}$ denote a \emph{Decision Tree}, i.e. a binary rooted tree with a `Decision Rule' (see next bullet point) associated to each node;
	\item A \emph{Decision Rule} is a pair $(n',p')$, with $n'\in \{ 1, \ldots, n\} $ and $p' \in \{1, \ldots, p \} $, corresponding to the partitioning of $\mathbb{R}^{n}$ induced by the value of observation $n'$ at variable $p'$, i.e. the partitioning of $\mathbb{R}^{n}$ into two subsets  $\{\textbf{X} \in \mathbb{R}^{p} : X_{p'} \leq X_{n', p'} \}$ and $\{\textbf{X} \in \mathbb{R}^{p} : X_{p'} > X_{n', p'} \}$. A decision rule cannot produce empty cells or contradict any Decision Rules from its ancestor nodes;
	\item  Let  $\textbf{A}_{k}= \{A_{k,1},\ldots,A_{k,L_{k}} \}$ denote as the corresponding partition of $\mathbb{R}^{p}$ induced by $\mathcal{T}_{k}$ and $\textbf{X}$,  where $L_{k}$ denotes the number of leaves in the $k$-th tree $\mathcal{T}_{k}$;
	\item  $\pmb{\mu}_{k}= \{\mu_{k,1},\ldots,\mu_{k,L_{k}} \}$ denotes the vector of leaf parameters, associated to each leaf node of tree $\mathcal{T}_{k}$.
\end{enumerate} 
With this notation, the $k$-th Regression Tree function in \eqref{eq:BART} (assuming a step function) is
\begin{equation*} 
	g(\textbf{X}_{i};\mathcal{T}_{k},\pmb{\mu}_{k}) = \sum_{j=1}^{L_{k}}\mu_{k,j} \mathbb{I}(\textbf{X}_{i} \in A_{k,j}).
\end{equation*}

\subsection{Indian Buffet Process} \label{sec:IBP}
The IBP is a distribution for binary matrices with a finite number of rows and an infinite number of columns.  It was first introduced in \citet{GG(05)} and widely applied for factor analysis in machine learning applications. In this subsection, we specifically describe the three-parameter version of the IBP introduced in \citet{TG(09)}. 

In Bayesian non-parametric setting, the IBP can be used as a prior distribution, $\pi(\textbf{W}_{n})$, for random binary matrices $\textbf{W}_{n}$ with $n$ rows and infinite number of columns, i.e. the state space of  $\textbf{W}_{n}$ is $\{0,1\}^{n\times \infty}$. A sample from $\pi(\textbf{W}_{n})$ has, with probability one, only finitely many non-zero columns, with all remaining columns consisting entirely of zeros. The three parameter version of the IBP has three hyper-parameters, denoted as $(\gamma,\delta, \eta)$. The parameter $\gamma>0$ regulates the total number of active columns (i.e. columns containing at least one $1$); larger values of $\gamma$ lead to more active columns.
$\eta \in ({-}\infty,1)$ regulates the proportion of $1$s and $0$s in each column, with smaller values of 
$\eta$ resulting in a higher proportion of $1$s in active columns. Finally, $\delta \in (-\eta, \infty)$ governs the amount of sharing of $1$s values across columns, thereby controlling the sparsity of the matrix: values of $\delta$ close to $-\eta$ produce similar columns, while larger values of $\delta+\eta$ (i.e., $\delta$ far from $-\eta$) generate sparse matrices, where columns have $1$s in different row positions.

Let $W_{i,\cdot}$ be the $i$-th row of $\textbf{W}_{n}$. The distribution $\pi(\textbf{W}_{n})$ can be specified by defining the marginal distribution of the first row, $\pi(W_{1,\cdot})$ and the conditional distribution $\pi(W_{i,\cdot}|W_{1,\cdot},\ldots,W_{i-1,\cdot})$ of the $i$-th row given the previous ones, as expressed by
$$\pi(\textbf{W}_{n}) = \pi(W_{1,\cdot})\pi(W_{2,\cdot}|W_{1,\cdot})\cdots \pi(W_{n,\cdot}|W_{1,\cdot},\ldots,W_{n-1,\cdot}).$$
Under an IBP prior for $\pi(\textbf{W}_{n})$, the corresponding marginal and conditional distributions are given by:
\begin{enumerate}
	\item $\pi(W_{1,\cdot})$: Sample $K_{1}\sim \text{Poisson}(\gamma)$, and set the first $K_{1}$ entries of $W_{1,\cdot}$ to $1$ and all remaining entries to $0$; that is, $W_{1,j}=1$ for $j\leq K_{1}$, and $W_{1,j}=0$ otherwise.
	\item $\pi(W_{i+1,\cdot}|W_{1,\cdot},\ldots,W_{i,\cdot})$, for $i = 1,\ldots, n-1$: Given $W_{1,\cdot},\ldots,W_{i,\cdot}$,  $W_{i+1,\cdot}$ is sampled in the following two steps: 
	\begin{enumerate}
		\item Let $K_{i}$ denote the number of columns containing at least one $1$ among the first $i$ rows, that is, all columns $k$ such that $\sum_{j=1}^{i}W_{j,k}>0$. The first $K_{i}$ entries of row $i+1$ are sampled from
		\begin{equation} \label{eq:prob.old}
			\mathbb{P}(W_{i+1,k}=1|W_{1,\cdot},\ldots,W_{i,\cdot}) =\frac{m_{k}-\eta}{i +\delta},
		\end{equation}
		where $m_{k}:=\sum_{j=1}^{i}W_{j,k}$ and is the number of $1$s in column $k$ up to row $i$ (for ease of notation, we avoid the subscript $i$ in $m_{k}$).
		\item A random number $K^{\text{new}}_{i+1} \sim \text{Poisson}(\gamma\frac{\Gamma(1+\delta)\Gamma(i+\delta+\eta)}{\Gamma(i+1+\delta)\Gamma(\delta+\eta)})$  of $1$s is added to the $(i+1)$-th row, i.e.  $W_{i+1,k}=1$ for $k=K_{i}+1,\ldots,K_{i} +K^{\text{new}}_{i+1}$. Finally, all remaining entries in row  $i+1$ are set to $0$, i.e. $W_{i+1,k}=1$ for $k>K_{i}+K^{\text{new}}_{i+1}$.
	\end{enumerate}
\end{enumerate}
Under this construction, we can derive the distribution of $\pi(\textbf{W}_{n})$, given the hyper-parameters $(\gamma,\delta,\eta)$, commonly referred to as the \emph{Exchangeable Feature Probability Function}. This distribution depends on the realized value of $\textbf{W}_{n}$ only through the number of active columns $K_{n}$ and the columns counts $(m_{1},\ldots,m_{K_{n}})$, i.e. $m_{k}:=\sum_{j=1}^{n}W_{j,k}$ the number of ones in column $k$ among all $n$ rows, and can be written as  
\begin{equation}\label{eq:EFPF}
	\frac{1}{K_{n}!} \left( \frac{\gamma}{(1+\delta)_{n-1\uparrow}}\right)^{K_{n}}  \exp\left(-\gamma \sum_{i=1}^{n} \frac{(\eta+\delta)_{i-1\uparrow}}{(\delta +1)_{i-1\uparrow}} \right) \prod_{l=1}^{K_{n}} (1-\eta)_{m_{l}-1\uparrow} (\delta+\eta)_{n-m_{l}\uparrow}
\end{equation}
where  $(x)_{n\uparrow}:=x(x+1) \cdots(x+n-1)$ and is the rising factorial.

\section{The Infinite BART} \label{sec:InfiniteBART}

\subsection{Model and Priors} \label{sec:model}
In this section, we aim to bridge the BART model with an IBP prior, which allows for an infinite collection of potential weak learners while using a binary weight matrix $\textbf{W}$ to select a finite subset of them in the regression function. In addition, due to the inclusion of the matrix $\textbf{W}$, different observations may select different subsets of weak learners. The regression function of Infinite BART extends that of classic BART in Equation~\eqref{eq:regression}, by letting $K=\infty$ and adding the $n\times \infty$ binary weight matrix $\textbf{W}$, 
\begin{equation} \label{eq:BART-IBP}
	Y_{i} = \sum_{k=1}^{\infty} W_{i,k} g(\textbf{X}_{i}; \mathcal{T}_{k},\pmb{\mu}_{k}) + \epsilon_{i}.
\end{equation}
with $\epsilon_{i}\stackrel{\text{iid}}{\sim}\mathcal{N}(0,\sigma^{2})$. 

The classic BART model with $K$ trees is recovered as a special case by setting $W_{i,k}= 1$ for $k\leq K$ and $W_{i,k}= 0$ for $k>K$, for all $i=1,\ldots,n$. Moreover, the extension of classic BART with unknown number of trees $K$ (and no heterogeneity) is obtained by setting $-\eta$ and $\delta$ very large with almost the same values. In this special case, the probability~\eqref{eq:prob.old} becomes almost 1, hence all observations either use or do not use each tree. Finally, by choosing $-\eta$ and $\delta$ small and/or far apart, the model specification~\eqref{eq:BART-IBP} allows for (soft) clusters of observations to have heterogeneous regression functions. {Heterogeneity in the regression function can arise naturally when important variables are omitted from the set of explanatory variables. In particular, the omission of significant factor or integer-valued variables can produce clusters of responses characterized by different regression functions, corresponding to different levels of the factor variable. When multiple factor variables are unobserved, soft clusters of regression functions may emerge across different combinations of their underlying factor levels.}
 As an illustration, at the end of Section~\ref{exp:prediction}, we fit BART and Infinite IBP on datasets in which a significant factor variable has been removed from the set of explanatory variables, and test their predictive performances.

{We emphasize that this generalization of BART is distinct from the idea of approximating BART by a Gaussian process, as sketched in Section 5 of \cite{linero2017review}. In Infinite BART, for each sample size $n$, the number of active trees, $K_n$, is finite, unknown, and learned from the data. The number $K_n$ may increase with the sample size and can potentially diverge to infinity as $n\to\infty$. By contrast, the heuristic argument in \cite{linero2017review} considers the limit $K\to\infty$ for the number of trees and for fixed $n$, potentially under an appropriate scaling, and suggests that BART may then be approximated by a Gaussian process. This heuristic is motivated by analogous results in the frequentist literature relating Random Forests to kernel estimators and Gaussian processes; see, for example, \cite{scornet2016random}.
}

The unknown parameters of the model are the binary matrix $\textbf{W}$, the trees  $\mathcal{T}_{k}$ and associated leaf parameters $\pmb{\mu}_{k}$, and the error variance $\sigma^{2}$. We assign a three-parameter IBP prior to the weight matrix $\textbf{W}$, and for the remaining parameters we adopt the same priors used in the BART model of \citet{CMG(10)}. Specifically, all trees are assumed a priori independent, each following a binary-split branching-process prior in which the probability that a node at depth $d$ is non-terminal is given by $\alpha (1+d)^{-\beta}$, for some hyper-parameters $\alpha\in (0,1)$ and $\beta\in[0,\infty)$. Conditionally on the tree $\mathcal{T}_{k}$, the leaf parameters $\mu_{kj}$ are assumed independent with prior $\mathcal{N}(0,\sigma_{\mu}^{2})$. The error variance is assigned prior $\sigma^{2}\sim (\nu\lambda)/\chi^{2}_{\nu}$. In the simulations of Section~\ref{sec:applications}, the hyper-parameters $\alpha,\beta,\sigma^{2}_{\mu},\nu,\lambda$ are set to the default values recommended in \citet{CMG(10)}.

The shape of $\textbf{W}$ sampled from an IBP is sensitive to the choice of the hyper-parameters $(\gamma,\delta,\eta)$. For this reason, these three hyperparameters are assigned hyper-priors and are learned a posteriori within the MCMC sampler described in subsection~\ref{sec:mcmc}. Specifically, in the simulations of Section~\ref{sec:applications}, we use the following hyper-priors: $\gamma \sim \text{Gamma}(a_{\gamma},b_{\gamma})$;  since $\eta$ has support $(-\infty,1)$, we place a prior on $1-\eta \sim \text{Gamma}(a_{\eta},b_{\eta})$ prior; and because $\delta$ takes values in $(-\eta,\infty)$, we assign a conditional prior on  $\eta+\delta$, namely $\eta+\delta \sim \text{Gamma}(a_{\delta},b_{\delta})$.

\subsection{The MCMC Sampler} \label{sec:mcmc}

A schematic description of the MCMC sampler used to perform posterior inference on the model parameters is provided in Algorithm~\ref{algo1}. In the following, $\Theta = (\sigma^2, \gamma, \delta, \eta)$ denotes the vector of unknown hyper-parameters of the model. $\pmb{\mathcal{T}}_{-k}$ denotes the list of all currently active trees, excluding the $k$-th one. Similarly, $\textbf{W}_{-i}$ denotes the matrix $\textbf{W}$ without the $i$-th row. {In Algorithm~\ref{algo1}, we explicitly specify the variables being conditioned on at each step. However, for ease of notation, throughout the remainder of the paper we generically refer to these variables as `rest'. }
\begin{algorithm}[!t]
	\caption{The MCMC sampler for $\pi(\pmb{\mathcal{T}},\pmb{\mu},\textbf{W},\Theta | Y,\textbf{X})$}
	\label{algo1}
	\For{$m$ in 1:L (number of MCMC iterations)}{
		\For{$k$ in 1:$K_{n}$}{
			Update $\mathcal{T}_{k}|\pmb{\mathcal{T}}_{-k},\Theta,\textbf{W},\textbf{X},Y$\ as described in Step 1;\\
			Update $\mu_{k}|\pmb{\mu}_{-k},\mathcal{T}_{k},\Theta,\textbf{W},\textbf{X},Y$\ as described in Step 2;
		}
		\For{$i$ in 1:$n$}{
			Update  $\textbf{W}_{i,\cdot}|\textbf{W}_{-i}\pmb{\mathcal{T}},\pmb{\mu},\Theta,\textbf{X},Y$\ as described in Step 3;
		}
		Update $\Theta|\pmb{\mathcal{T}},\pmb{\mu},\textbf{W},\textbf{X},Y$\ as described in Step 4;
	}
\end{algorithm}

Specifically, each of the steps outlined in Algorithm~\ref{algo1} can be implemented as follows:
\begin{itemize}[leftmargin=*]
	\item \textbf{Step 1 (Update $\mathcal{T}_{k}$):} 
	The $k$-the tree can be updated using the backfitting algorithm described in \citet{CMG(10)}, which is also employed in generalized additive models \citep{HT(00)}. The only modification in our setting is that the update for tree $k$ uses only those observations $Y_{i}$ for which the tree is active, i.e., those satisfying $W_{i,k}=1$.  Therefore, we can define the $m_{k}$-dimensional vector, where  $m_{k}=\sum_{i=1}^{n} W_{i,k}$, of residuals $R^{(k)}$
	\begin{equation*}
		R^{(k)}_{i}=Y_{i}- \sum_{k' \neq k} W_{i,k'} g(\textbf{X}_{i}; \mathcal{T}_{k'},\mu_{k'})
	\end{equation*}
	for all $i=1,\ldots,n$ such that $W_{i,k}=1$. Given the $m_{k}$-vector of residuals, $\mathcal{T}_{k}|R^{(k)},\sigma^{2}$ is then updated using the Metropolis-Hastings (MH) proposal of \citet{CGM(98)}, which proposes a new tree configuration using one of the four possible moves: Grow, Prune, Change, Swap (see subsection 5.1 of \citet{CGM(98)} for more details).
	
	\item \textbf{Step 2 (Update $\mu_{k}$):} The update of $\mu_{k}$ is obtained from a conjugate Normal distribution, as in \citet{CGM(98)} and \citet{CMG(10)}. For the tree structure $\mathcal{T}_{k}$, the update follows the same procedure as in \citet{CMG(10)} with the only difference that it is based only on the $m_{k}$ observations currently assigned to tree $k$.
	
	\item \textbf{Step 3 (Update $W_{i,\cdot}$):} Due to the exchangeability of the rows of $\textbf{W}$, $W_{i,\cdot}$ can be treated as if it were the last row. Its update combines the IBP prior conditional on $\textbf{W}_{-i}$ (see bullet point 2 in subsection~\ref{sec:IBP}) with the likelihood in~\eqref{eq:BART-IBP}. After removing the zero columns in $\textbf{W}_{-i}$, the update $W_{i,\cdot}$ can be proceeded in two steps: 
	\begin{enumerate}[nosep]
		\item Let $K_{n}^{-i}$ denote the number of non-zero columns in $\textbf{W}_{-i}$, for $k=1,\ldots, K_{n}^{-i}$, the entry $W_{i,k} \in \left\{0,1\right\}$ is updated by sampling from $\mathbb{P}[ W_{ik} = w \,|\text{rest}]\propto$
		\begin{align*}
			& \mbox{ } 
			& \left\{\begin{array}{lll}
				\frac{m_{k}^{-i}-\eta}{i + \delta}\cdot \phi\left( Y_{i}; \sum_{k' \neq k} W_{i,k'} g(\textbf{X}_{i}; \mathcal{T}_{k'},\mu_{k'}) +g(\textbf{X}_{i}; \mathcal{T}_{k},\mu_{k}), \sigma^2 \right)   \hspace{0.2cm} &\text{for }w=1 \\[0.4cm]
				\frac{i + \delta - m_{k}^{-i}+\eta}{i + \delta} \cdot \phi \left( Y_{i}; \sum_{k' \neq k} W_{i,k'} g(\textbf{X}_{i}; \mathcal{T}_{k'},\mu_{k'}) , \sigma^2 \right) \hspace{0.2cm} &\text{for  } w=0
			\end{array}\right.
		\end{align*} 
		where $\phi(y;\mu,\sigma^2)$ denotes the density of a Normal distribution with mean $\mu$ and variance $\sigma^2$ evaluated at $y$, and $m_{k}^{-i} = \sum_{j\neq i}W_{j,k}$.
	
		\item A total of $K^{(i)}$ new trees are introduced, where $K^{(i)}$ is sampled from 
		\begin{align*}
			\mathbb{P}(K^{(i)}=k^{(i)}| & \text{rest}) \mbox{ }  \propto \mbox{ } \text{Poisson}\left(k^{(i)}; \gamma\frac{\Gamma(1+\delta)\Gamma(n-1+\delta+\eta)}{\Gamma(n+\delta)\Gamma(\delta+\eta)}\right) \times \\
			& \mbox{ } \mbox{ } \mbox{ }  \phi \left( Y_{i}; \sum_{k' =1}^{K_{n}^{-i}} W_{i,k'} g(\textbf{X}_{i}; \mathcal{T}_{k'},\mu_{k'}) + \sum_{k=1}^{k^{(i)}}g(\textbf{X}_{i}; \mathcal{T}^{(i)}_{k},\mu^{(i)}_{k}), \sigma^2 \right).   
		\end{align*}
		A total of $k^{(i)}$ columns are added in $\mathbf{W}$, each initialized with all entries equal to $0$ except for a single $1$ in row $i$. Additionally, $k^{(i)}$ new trees and their associated leaf parameters, $\mathcal{T}^{(i)}_{k},\mu^{(i)}_{k}$, are introduced and sampled following Step 1 and 2, using only the observation $Y_{i}$.
	\end{enumerate}
	\item \textbf{Step 4 (Update $\Theta$)}: The full conditional distributions of $\sigma^{2}$ and $\gamma$ are available in closed form, following an Inverse-Gamma and a Gamma distribution (using \eqref{eq:EFPF} with the prior for $\gamma$). In particular, $\gamma |\text{rest} \sim \text{Ga}(K_{n}+a_{\gamma}, \sum_{i=1}^{n} \frac{(\eta+\delta)_{i-1\uparrow}}{(\delta +1)_{i-1\uparrow}} +b_{\gamma})$.  {Specifically, we jointly update $\eta$ and $\delta$ using slice sampling \citep{N(03)}. To facilitate sampling under the constraints $\eta<1$ and $\eta+\delta>0$, we reparameterize those parameters as $u_1=\eta, u_2=\log(\eta+\delta)$. Conditional on other parameters, we update $u_1$ and $u_2$ sequentially using univariate slice sampling. For each parameter, an auxiliary slice level is first sampled from the exponential distribution on the log-posterior scale. An initial interval of width is then expanded using the stepping-out procedure, with at most $m$ expansions, followed by uniform sampling within the interval and shrinkage until a point within the slice is obtained.}
\end{itemize}

\paragraph{Initialization of the hyperparameters $(\gamma,\delta,\eta)$}
As mentioned previously, the shape of $\mathbf{W}$ depends on the hyperparameters  $(\gamma,\delta,\eta)$ of the IBP prior. These hyper-parameters should be randomized with hyper-priors and learned from the data.  A default initialization in the MCMC sampler is $(\gamma,\delta,\eta) = (-100,101,100)$, yielding a dense $\mathbf{W}$, closely mimicking a standard BART model. However, if the posterior of $\mathbf{W}$ favors much sparser and/or with fewer columns matrices, a long burn-in might be needed for the chain to leave this poor initialization. 
Alternatively, the starting values of $(\gamma,\delta,\eta)$ can be selected by performing short pilot runs (say 500 iterations), for a small set of candidate values.
If, in a pilot run, the number of active trees  mixes well, the final values of $(\gamma,\delta,\eta)$ from that pilot run can be used as starting values for the full MCMC sampler.
Overall, the output of the MCMC does not depend on the initial values of the chain, provided that  the algorithm is run for an adequate number of iterations. Nevertheless, a good initialization can  improve computational efficiency by shortening the required burn-in.

\subsection{Some statistical tasks} \label{sec:tasks}

The Infinite BART model can be used to solve the same statistical tasks as BART, with only minor modifications. In this subsection, we outline several of these tasks, while illustrative applications are deferred to Section~\ref{sec:applications}.

\paragraph{Point and interval estimation of $f(x)$:} The model can be used to infer the unknown regression function $f$. Specifically, it is possible to perform point estimation for the regression function $f(x)$, both for in-sample and out-of-sample values of $x$ by averaging over posterior draws. For an in-sample point $x$, an estimator of $f(x)$ is
\begin{equation} \label{eq:est.f(x)}
	f^{*}(x)=  \frac{1}{L}\sum_{l=1}^{L}\sum_{k=1}^{K_{n}^{(l)}}W^{(l)}_{\cdot,k}g(x;\mathcal{T}_{k}^{(l)},\mu^{(l)}_{k}),
\end{equation}
where $L$ is the number of iterations of MCMC (after discarding burn-in, and possibly applying thinning). For out-of-sample $x$, $W^{(l)}_{\cdot,k}$ in \eqref{eq:est.f(x)} is replaced by $\mathbb{E}(W^{(l)}_{\cdot,k}|\text{rest})=\mathbb{P}(W^{(l)}_{\cdot,k}=1|\text{rest})$, which can be computed from \eqref{eq:prob.old} using the posterior sample of $(\gamma^{(l)},\delta^{(l)},\eta^{(l)})$ and the corresponding posterior counts $m_{k}^{(l)}$ from the $l$-th MCMC iteration. Alternatively, it can be approximated through an additional Monte Carlo step, similar to  the procedure used for prediction $\mathbf{Y}$. Credible posterior regions for $f(x)$ can be derived in a similar manner, using empirical quantiles from the MCMC output.

\paragraph{Prediction:}
The model can predict $N$ new responses, $\textbf{Y}^{\text{pred}} := Y_{(n+1):(n+N)}$, within the MCMC sampler, given additional values of the explanatory variables, $\textbf{X}_{(n+1):(n+N),\cdot} \in \mathbb{R}^{N\times p}$. To predict $\textbf{Y}^{\text{pred}}$, at the end of each MCMC iteration $l$, the sampler draws $\mathbf{W}^{(l)}_{(n+1):(n+N)}$ from 
$\pi(\mathbf{W}^{(l)}_{(n+1):(n+N)}|\hat{Y}^{(l-1)}_{(n+1):(n+N)},\text{rest})$, using the same update for $\mathbf{W}$ described in subsection~\ref{sec:mcmc}. It then sets
\begin{equation*}    \hat{Y}^{(l)}_{i}=\sum_{k=1}^{K^{(l)}_{n+N}}W^{(l)}_{i,k}g(\textbf{X}_{i};\mathcal{T}^{(l)}_{k},\mu^{(l)}_{k}),
\end{equation*}
for all $i=n+1,\ldots,n+N$. An alternative approach, used in the simulations, is to initialize $\mathbf{W}^{(l)}_{(n+1):(n+N)}$ from the IBP conditional on $\mathbf{W}^{(l)}_{1:n}$ and $(\gamma^{(l)},\delta^{(l)},\eta^{(l)})$. After that,  $\hat{Y}^{(l)}_{(n+1):(n+N)}$ is generated by alternately sampling from $\pi(Y_{(n+1):(n+N)}|\text{rest})$ and $\pi(\mathbf{W}_{(n+1):(n+N)}|\text{rest})$ {$M$} times, retaining the final draw of $Y$ as $\hat{Y}^{(l)}_{(n+1):(n+N)}$. {As default value, used in all simulations, we set $M=10$.} This approach has the advantage that the prediction step needs to be performed only at thinned iterations.

\paragraph{Variable selection:}
As in the classic BART model, the importance of each explanatory variable can be estimated by computing
\begin{equation} \label{eq:var.imp}
	v_{j} = \frac{1}{L}\sum_{l=1}^{L} z^{(l)}_{j},
\end{equation}
for $j=1,\ldots,p$, where $z^{(l)}_{j}$ is the proportion of splitting rules using variable $j$, among all $K^{(l)}_{n}$ trees that are active at iteration $l$, $(\mathcal{T}^{(l)}_{1},\ldots,\mathcal{T}^{(l)}_{K^{(l)}_{n}})$. 

In the experiments of Section~\ref{sec:applications}, we use the standard Variable Importance metric \eqref{eq:var.imp}, which is used also in \cite{CMG(10)}. However, we remark that alternative approaches to perform variable selection with Bayesian tree methods have been proposed in the literature, such as the Dirichlet prior of  \citep{L(18)}, the method proposed by \cite{bleich2014variable} or the Metropolis-Hastings information metric of \cite{LuoDaniels2024}. 

\paragraph{Estimation of partial dependence function:} The partial dependence function \citep{F(01)} summarizes the marginal effects of one or more explanatory variables on the response variable.  Specifically,  the partial dependence  function of the explanatory variable $x_s$ on $Y$ is defined as 
\begin{equation} \label{eq:partial}
	f(x_{s}) := \frac{1}{n}\sum_{i=1}^{n} f(x_{s},\textbf{X}_{i}^{-s}),
\end{equation}
where the vector $\textbf{X}$ has been partitioned into the variable of interest, $x_{s}$, and the remaining $p-1$ explanatory variables $\textbf{X}^{-s}$.
Estimation of such a quantity can be performed in a similar way as in BART. Specifically, for a grid of values for $x_{s}$, we can estimate $\widehat{f(x_{s})}= \frac{1}{n}\sum_{i=1}^{n}f^{*}(x_{s},\textbf{X}_{i}^{-s})$, where $f^{*}$ is defined in \eqref{eq:est.f(x)}.

\paragraph{Estimation of casual effects:} {Although our primary focus is predictive modelling, the proposed framework may also have potential applications in causal inference as in \cite{H(11)}. BART has become an important tool for flexible causal modelling because its nonparametric regression structure can accommodate nonlinearities and interactions when estimating heterogeneous treatment effects. For example, Bayesian causal forests introduce a BART-based framework specifically designed to estimate heterogeneous treatment effects while addressing regularization and confounding by observed covariates \citep{HR(20)}, with subsequent extensions introducing additional shrinkage to improve performance in settings with many potentially irrelevant covariates \citep{caron2022shrinkage}. More broadly, recent work has developed a range of machine learning approaches \citep{chernozhukov2018double, hu2023heterogeneous} and Bayesian nonparametric methods \citep{LY(23)} for heterogeneous treatment effects. To illustrate how the proposed framework could be used in this setting,} let us denote $T\in\{0,1\}$ as the treatment variable, $Y$ as the outcome and $\textbf{X}$ as a set of confounders. We can estimate the ATE by evaluating the difference of the regression function at $T=1$ and $T=0$ using the posterior predictive. That is, for each posterior draw, we canhe individual treatment effect for unit  $i$ as $\tau_i = f^{*}(1,\mathbf{X}_{i}) - f^{*}(0,\mathbf{X}_{i})$, and then average over all subjects to obtain one draw of the posterior predictive ATE.  Repeating this across all posterior draws provides the posterior distribution of the ATE. {This illustrates how the flexible heterogeneous regression structure of the proposed method could naturally be extended to causal estimands.}


\section{Experiments}\label{sec:applications}

In this section, we test the performance of Infinite BART on both simulated and real data. Specifically, in subsection~\ref{exp:friedman}, we test the capabilities of Infinite BART to perform variable selection and estimate the partial dependence function. {In subsection~\ref{sec:timing}, we compare the computational time of classic and infinite BART.} In subsection~\ref{exp:prediction}, we compare the out-of-sample predictive performances of Infinite and classic BART in different simulated and real data scenarios. Finally, subsection~\ref{exp:casual} provides an illustration of casual effects estimation with Infinite BART. 

In all experiments, the hyper-parameters in BART are set to the following default values, consistent with those recommended in \citet{CMG(10)}, for both the classic and Infinite BART models: $\alpha=0.95$, $\beta=2$. {Also, in BART, we set the number of trees to $K=200$, and $\sigma^{2}\sim \nu \lambda / \chi ^{2}_{\nu}$, with $\lambda$ chosen in the data dependent manner (with $\nu=3$ and $q=0.9$), as recommended in subsections 2.2.4 and 2.2.5 of \citet{CMG(10)}.}  {For $\sigma^{2}_{\mu}$, in Infinite BART, we first conduct a sensitivity analysis using Friedman's example as in Section \ref{exp:friedman}. We consider $\sigma_\mu^2 \in \{0.01, 0.1, 0.5, 1, 10\}$, with five repetitions for each setting. The corresponding MSEs are 8.08, 4.40, 5.47, 5.28, and 5.11, respectively. Therefore, we use $\sigma_\mu^2=0.1$ as default value for all subsequent analyses. Alternatively, $\sigma_\mu^2$ can be updated within the sampler using Metropolis–Hastings step (this option is implemented in the code). As for BART, we follow \cite{CMG(10)} recommendation to set $\sigma_\mu^2$ proportional to the inverse of the number of trees, specifically $\sigma^2_{\mu} =(0.5/(2\sqrt{K}))^2$, and  keep $\sigma_\mu^2$ fixed to this value throughout our experiments.} The prior hyper-parameters for $(\gamma,\delta,\eta)$ are chosen to yield large prior variances, following the reparameterization described in subsection~\ref{sec:IBP}: $a_{\eta} = 0.05$, $b_{\eta} = 0.01$, $a_{\delta} = 0.1$, $b_{\delta}= 0.01$, $a_{\gamma}= 0.05$, $b_{\gamma} = 0.01$. 

The classic BART is implemented using the code supplied in \citet{SPI(23)}, publicly available in the two Github folders: \href{https://github.com/ebprado/AMBARTI}{https://github.com/ebprado/AMBARTI} and \href{https://github.com/ebprado/ExtensionsBART}{https://github.com/ebprado/ExtensionsBART}.  The Infinite BART implementation used in the experiments is a modification of this code, available at

 \href{https://github.com/yumcgill/Infinite-BART}{https://github.com/yumcgill/Infinite-BART}. 

This allows for a fair comparison of the two models’ performance, independent of differences in their implementations. {Also, the Infinite BART implementation in Github also includes the extension for classification, using the Probit link function and data augmentation scheme of \cite{AC(93)}, following the same extension for BART described in \cite{CMG(10)}.}

		\subsection{Variable Selection and Partial Dependence} \label{exp:friedman}
		
		In this subsection, we evaluate Infinite BART for variable selection and partial dependence function estimation using the benchmark Friedman regression function. Specifically, for each observation, $p$ explanatory variables are simulated from a Uniform distribution, i.e. $X_{i,1},\ldots,X_{i,p} \stackrel{\text{iid}}{\sim}\text{Unif}(0,1)$. The response variable $Y_{i}$ is simulated from
		\begin{equation} \label{eq:Friedman}
			Y_{i} = 10\cdot\text{sin}(\pi x_{i,1}x_{i,2})+ 20(x_{i,3} -0.5)^{2}+10x_{i,4} +5x_{i,5}+ \epsilon_{i}
		\end{equation}
		where $\epsilon_{i}\sim N(0,1)$, for $i=1,\ldots,n$. This example features a true regression function that includes both linear and nonlinear components, an interaction term, and irrelevant variables when $p>5$.
		
\begin{figure}
			\centering
			\includegraphics[width=1\linewidth]{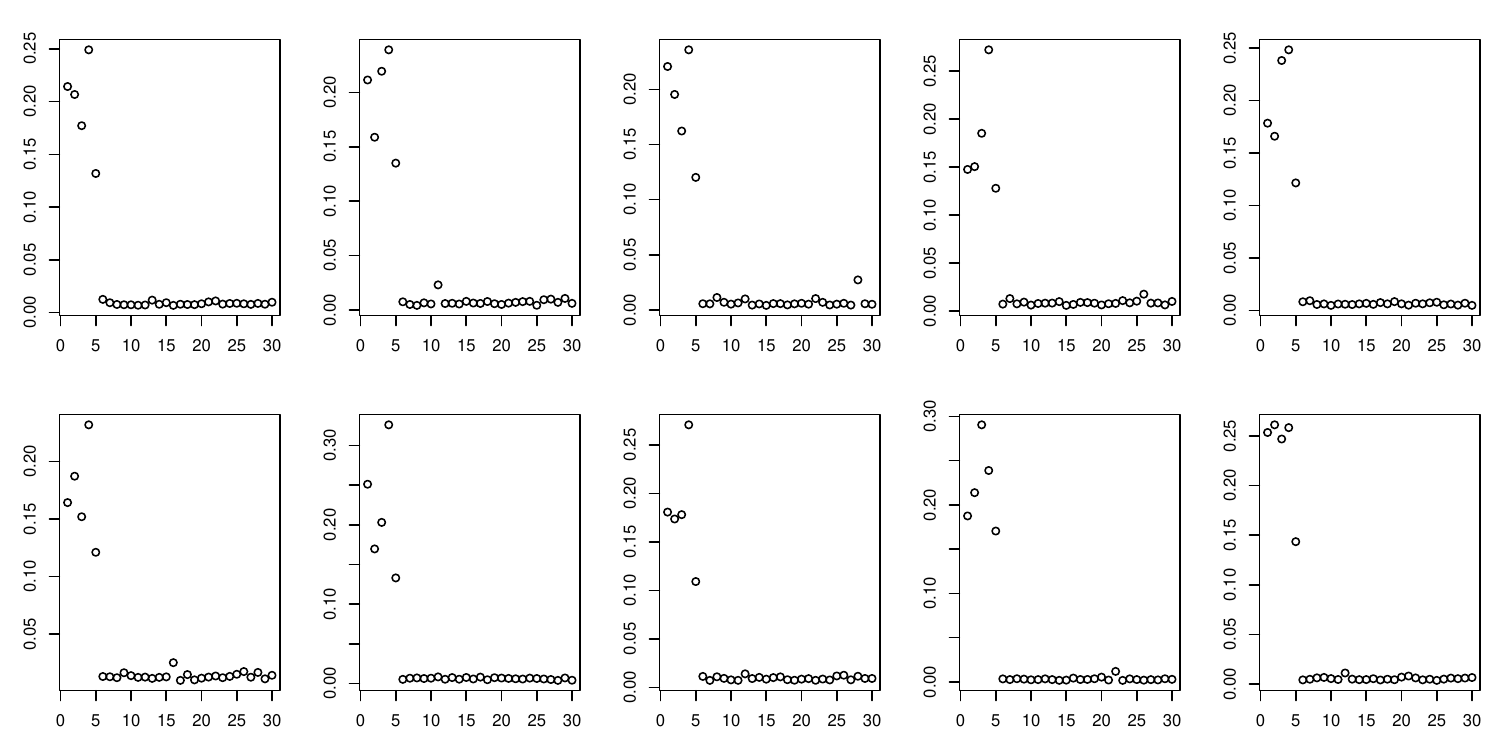}
			\caption{Important variables for Friedman's example of ten replicates.}
			\label{fig:fried.var.imp}
		\end{figure}

		We generated 9 synthetic datasets with $n=300$ and $p=30$ according to \eqref{eq:Friedman}. For each dataset, the Infinite BART model was fitted for 5,000 iterations, following a burn-in of 1,000 iterations and without any thinning. Figure~\ref{fig:fried.var.imp} shows the estimated variable importance (formula~\eqref{eq:var.imp}) for all 30 variables across the 9 runs. In all cases, the model successfully identifies the first five variables as important, while assigning near-zero importance to the remaining 25 irrelevant variables that are not included in \eqref{eq:Friedman}.

		A comparison with Figure 5 of \citet{CMG(10)} might be relevant. That figure shows that in the classic BART model, variable importance estimates are highly sensitive to the number of trees. When a large number of trees is used (e.g., 100 or 200, which are default values in most BART packages), the estimates deteriorate: irrelevant variables may receive importance scores comparable to those of the true predictors, even with only $p=10$ variables. On the contrary, using only 10 trees yields very accurate estimates, correctly identifying the important variables. However, in real-world applications, the true important variables are unknown, making it difficult to choose the appropriate number of trees for variable selection. 	Instead, the Infinite BART model automatically selects the number of trees from the data. In the 9 replicates of the Friedman example, the estimated number of active trees were {$10.03, 12.98, 13.22, 16.60, 23.85, 14.03,  8.01, 11.024,  6.11,  8.03$}, which explains the high accuracy of variable importance in Figure~\ref{fig:fried.var.imp}.

\begin{figure}
			\centering
			\includegraphics[height=8cm,width=1\linewidth]{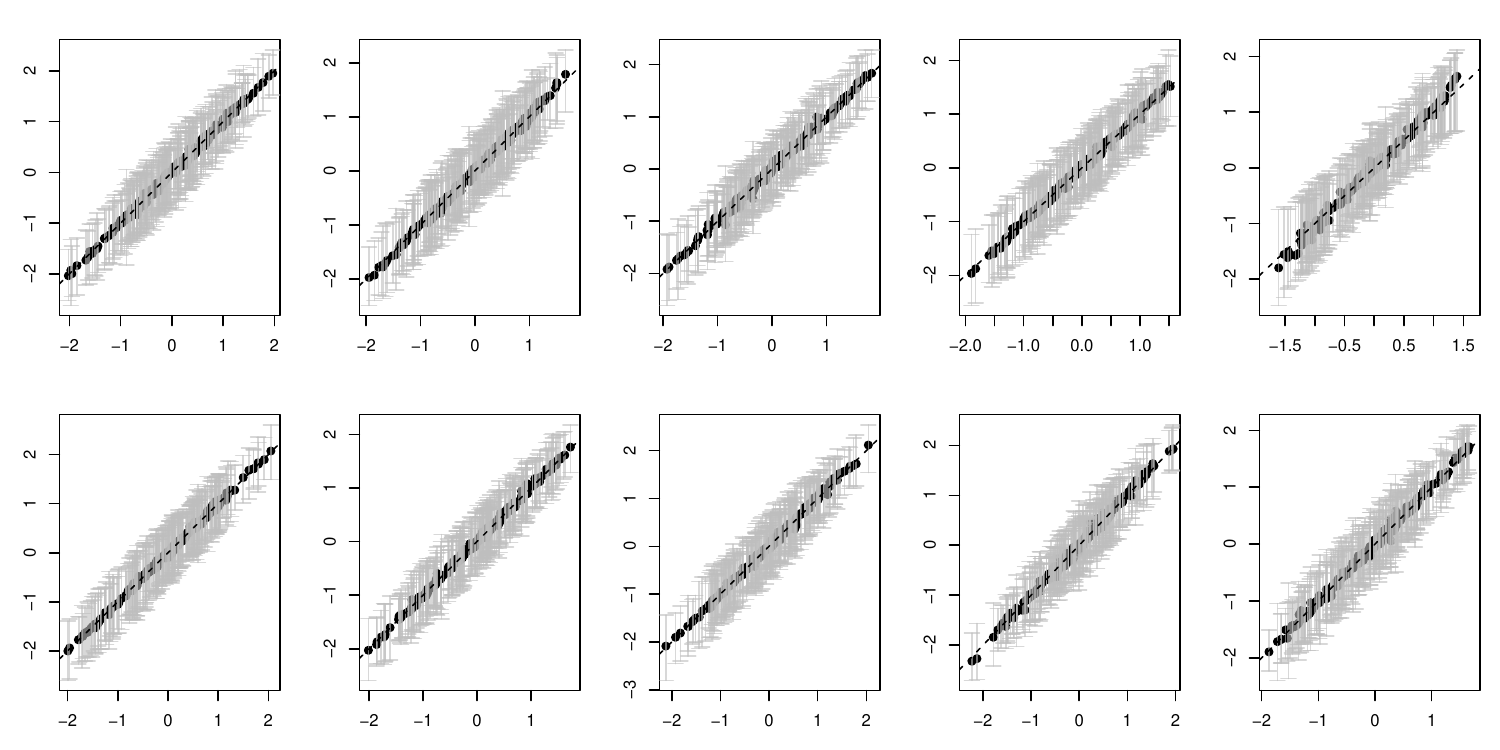}
			\caption{In sample inference about Friedman’s example of ten replicates.}
			\label{fig:fried.pred.fx}
		\end{figure}

\begin{figure}
			\centering
			\includegraphics[height=6cm,width=1\linewidth]{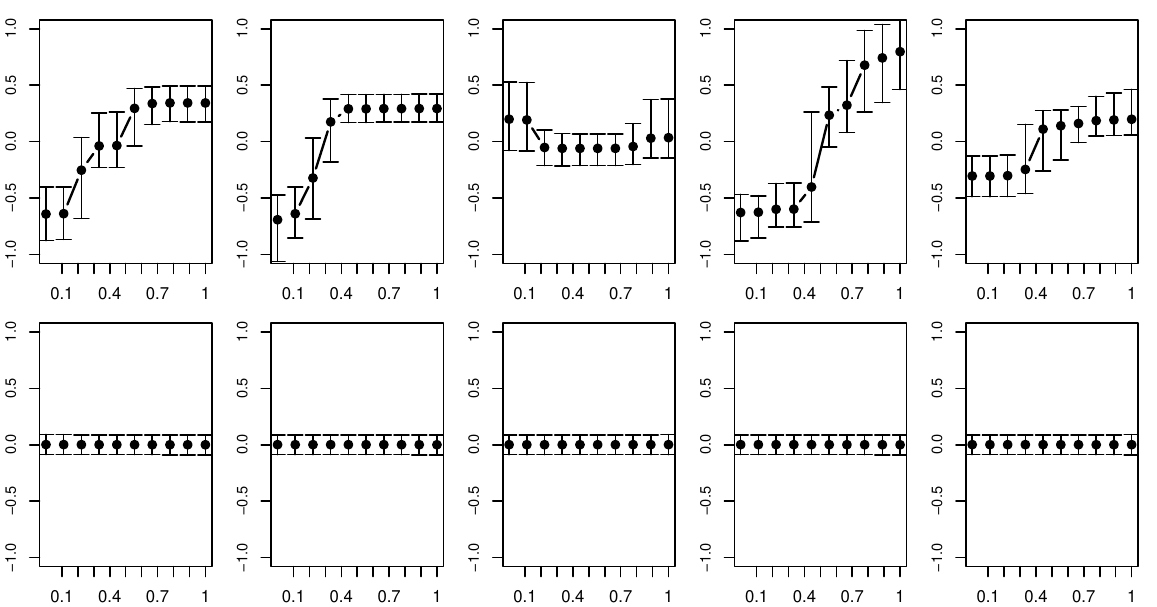}
			\caption{Partial dependence for the first 10 predictors in the Friedman data, one replicate.}
			\label{fig:fried.part.dep}
		\end{figure}

		Figures~\ref{fig:fried.pred.fx} and~\ref{fig:fried.part.dep} display the in-sample predictions and partial dependence function estimates. Specifically, Figure~\ref{fig:fried.pred.fx} plots the estimated values of $f^{*}(\mathbf{X}_{i})$, using formula~\eqref{eq:est.f(x)}, against the true values of $Y_{i}$, including 95\% credible regions for $f(\mathbf{X}_{i})$. Across all replicates, the fitted values closely match the true values, and their credible intervals have reasonable width. Figure~\ref{fig:fried.part.dep} shows the estimated partial dependence function (formula~\eqref{eq:partial}) for the first run and the first 10 explanatory variables. The plots (not shown) of the remaining 20 variables are nearly identical to those of the 5 irrelevant variables (variables 6-10) shown in the bottom row. The partial dependence estimates capture the true functional relationships: variables 1 and 2 have identical effects, reflecting their identical contribution in \eqref{eq:Friedman}; variable 3 decreases before 0.5 and increases afterward, consistent with \eqref{eq:Friedman}; variables 4 and 5 plots are both strictly increasing, with the slope of variable 4 roughly twice that of variable 5; all other variables show flat, near-zero dependence on the response.

\begin{figure}
			\centering
			\includegraphics[height=6cm,width=1\linewidth]{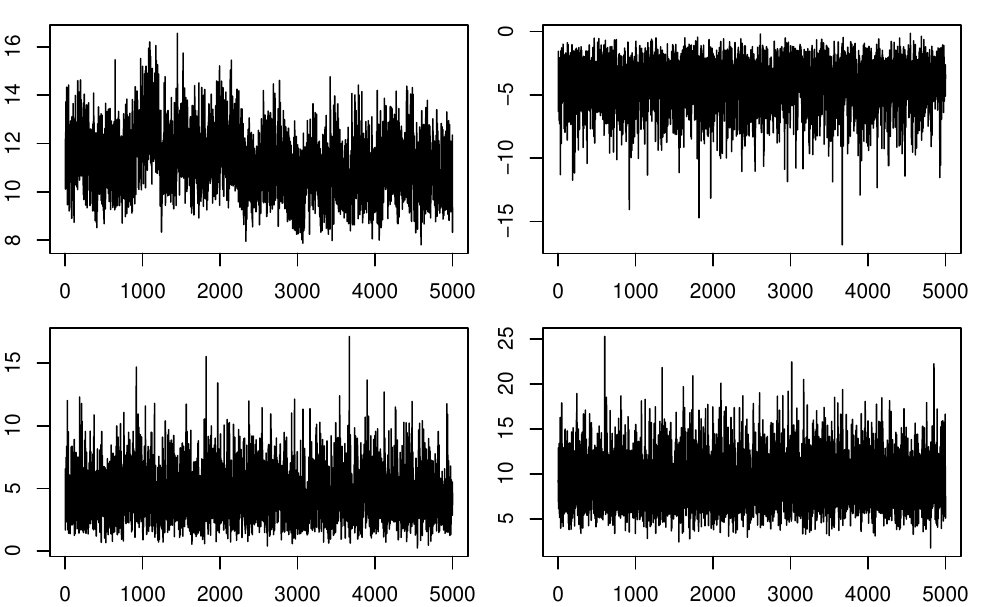}
			\caption{Trace plots in the Friedman data: top left $\sigma^2$; top right $\eta$; bottom left $\delta$; bottom right $\gamma$.}
			\label{fig:fried.hyper}
		\end{figure}

		Finally, Figure~\ref{fig:fried.hyper} shows the trace plots of the hyperparameters $\Theta = (\sigma^2, \gamma, \delta, \eta)$, which indicate reasonably good mixing (without any thinning) using slice sampling and full conditional updates.
		
			\subsection{Computational cost of BART and Infinite BART}
			\label{sec:timing}
			We compare the computing time of the MCMC samplers of BART \citep{CMG(10)} and Infinite BART.
			Data are simulated from the benchmark Friedman function of \citet{F(01)}, formula \eqref{eq:Friedman},
			with $\varepsilon_i \sim N(0,1)$ and $x_i \sim U(0,1)^p$ independently, for sample sizes
			$n \in \{ 100, 500, 1000, 5000, 10.000 \}$, and numbers of covariates $p \in \{ 10, 50, 100, 200\}$. For each $(n,p)$, both samplers are run on the same data set for the same number of iterations
			($200$ iterations: $100$ burn-in and $100$ further iterations), and we record the elapsed time of each run,
			including initialisation. BART uses $K = 200$ trees. Infinite IBP is initialised so that a priori each observation
			uses $40$ trees, two observations share a given tree with probability $0.5$, and the expected total
			number of trees is $100$. In the runs, Infinite BART used on average between $76$ and $138$ trees in total.
			Both models use the prior calibration of \citet{CMG(10)} on the standardised response,
			$\alpha = 0.95$, $\beta = 2$ and uniform selection of the splitting variables. Both samplers share the
			same implementation of the tree moves, and all runs used a single core of the same machine.

		\begin{table}[ht]
			\centering
				\small
				\setlength{\tabcolsep}{5pt}
				\renewcommand{\arraystretch}{1.05}
				\begin{tabular}{rccc@{\hspace{1.2em}}rccc}
					\toprule
					\multicolumn{4}{c}{$p=10$} &
					\multicolumn{4}{c}{$p=50$} \\
					\cmidrule(lr){1-4}\cmidrule(lr){5-8}
					$n$ & BART & Infinite BART & Ratio &
					$n$ & BART & Infinite BART & Ratio \\
					\midrule
					100    & 5.8  & 4.9  & 0.85 & 100    & 5.4  & 3.9  & 0.73 \\
					500    & 6.3  & 5.9  & 0.93 & 500    & 6.4  & 5.7  & 0.90 \\
					1,000  & 7.8  & 9.7  & 1.24 & 1,000  & 7.7  & 10.0 & 1.30 \\
					5,000  & 19.4 & 34.0 & 1.75 & 5,000  & 19.9 & 41.2 & 2.07 \\
					10,000 & 33.0 & 67.6 & 2.05 & 10,000 & 33.5 & 72.9 & 2.18 \\
					\midrule
					\multicolumn{4}{c}{$p=100$} &
					\multicolumn{4}{c}{$p=200$} \\
					\cmidrule(lr){1-4}\cmidrule(lr){5-8}
					$n$ & BART & Infinite BART & Ratio &
					$n$ & BART & Infinite BART & Ratio \\
					\midrule
					100    & 5.1  & 5.5  & 1.07 & 100    & 5.4  & 3.8  & 0.71 \\
					500    & 6.1  & 7.6  & 1.25 & 500    & 6.3  & 8.7  & 1.38 \\
					1,000  & 7.5  & 9.8  & 1.30 & 1,000  & 7.5  & 12.0 & 1.60 \\
					5,000  & 19.5 & 43.5 & 2.23 & 5,000  & 19.5 & 50.3 & 2.58 \\
					10,000 & 33.2 & 67.8 & 2.04 & 10,000 & 34.2 & 97.0 & 2.84 \\
					\bottomrule
				\end{tabular}
				\caption{Computing time of the MCMC, in seconds per $100$ iterations
					(single core), and ratio of the times (Infinite BART / BART).}
				\label{tab:timing}
		
		\end{table}
		
	Table~\ref{tab:timing} shows that the computing time of BART is almost unaffected by the number of
			covariates, and grows by a factor of about $6$ between $n = 100$ and $n = 10{,}000$
			(from about $5$--$6$ to about $33$--$34$ seconds per $100$ iterations), less than proportionally to
			$n$ because a large part of its cost is a fixed overhead per tree move. For small samples Infinite IBP is faster than BART in three of the four settings (with $n = 100$: ratios in $0.71$--$0.85$, expect $p = 100$ with ratio $1.07$), since each observation uses only part of the trees.
			Its cost, however, grows faster with $n$, as the update of the binary matrix $W$ visits all
			$n \times K_{n}$ entries at every iteration, with $K_{n}$ being the current number of trees. Infinite BART is about as fast as BART for $n = 500$ when $p \leq 50$, and $1.2$--$1.6$ times slower for $n = 1000$, $1.7$--$2.6$ times for $n = 5000$ and $2.0$--$2.8$ times for $n = 10.000$. In general, infinite BART scales worse than BART with respect to the sample size. 
			For instance, $2000$ iterations in the worst scenario, at $n = 10.000$ and $p = 200$,  takes about one third of the time for BART to run compared to Infinite BART, on a single core.
			For large $n$, the cost of Infinite BART seems to increase also with $p$ (from $68$ to $97$ seconds, per $100$ iterations at $n = 10.000$, from $p=10$ to $p=200$, respectively), even though not too dramatically. Times scale linearly with the number of iterations for both algorithms.

		\subsection{Out-of-sample Prediction}~\label{exp:prediction}
		To evaluate the out-of-sample performance of Infinite BART, we compare it with classic BART in three settings: simulated data, real data, and `clustered' data. For each dataset (synthetic or real), we generate 10 independent train-test splits, using 4/5 of the data for training and 1/5 for testing, and assess prediction accuracy using the mean squared error.
		
		\paragraph{Simulated Data:}
		10 samples of synthetic data of size $n=100$ were simulated from the model for different values of  $(\gamma,\delta,\eta)$. These hyper-parameters control the number of trees and the sparsity of the matrix $\mathbf{W}$ (see subsection~\ref{sec:IBP}). Specifically, when $-\eta$ and $\delta$ are both large and similar in value (so that $\delta+\eta$ is small), the resulting matrix $\mathbf{W}$ is very dense, with most entries equal to $1$ (the probability in formula~\eqref{eq:prob.old} approaches $1$). Conversely, small values of $-\eta$ combined with large values of $\delta+\eta$ yield sparse matrices, with reduced sharing of trees among observations.
		The parameter $\gamma$ regulates the overall number of trees as each observation has a marginal number of active trees distributed as $\text{Poisson}(\gamma)$. 
		
		Figure~\ref{tab:mse.sim} displays the MSEs for each of the 10 runs and Table~\ref{tab:mse.sim} reports the corresponding average MSEs for the different choices of hyperparameters. For $(\gamma,\delta,\eta)=(100,101,-100)$ (or $(10,101,-100)$), the resulting $\mathbf{W}$ matrix has around $100$ (or $10$) columns, with all entries equal to $1$. Hence, it closely resembles the structure in BART, and the MSEs of classic and Infinite BART are similar. In contrast, for larger values of of $\delta+\eta$, the matrix $\mathbf{W}$ becomes sparse, introducing more heterogeneity among observations, and the MSE of BART tends to be consistently higher than that of Infinite BART.
		
		\begin{figure}
			\centering   \includegraphics[height=7cm,width=1\linewidth]{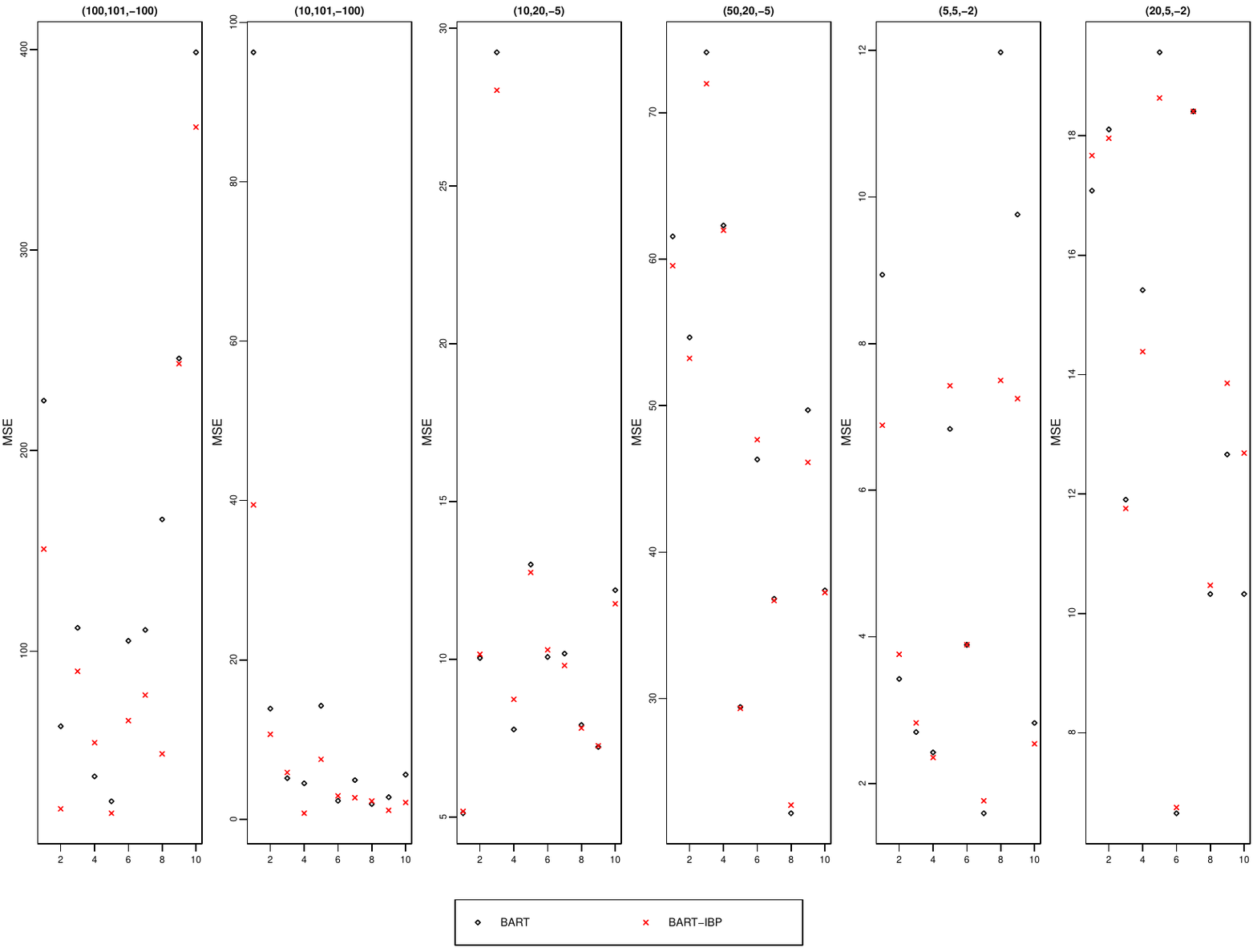}
			\caption{Synthetic data from the model. MSEs on 10 runs for different values of $(\gamma,\delta,\eta)$. Black dots for BART; red crosses for Infinite BART.}
			\label{fig:mse.sim}
		\end{figure}
		
		\begin{table}[ht]
			\centering
			\begin{tabular*}{28pc}{@{\hskip5pt}@{\extracolsep{\fill}}c@{}c@{}c@{}c@{\hskip5pt}}
				\toprule
				$(\gamma,\delta,\eta)$ & $(100,101,-100)$ & $(10,101,-100)$ & $(10,20,-5)$   \\
				\midrule
				BART & {148.8160}& {15.1512} & {11.2757}   \\
				Infinite BART &  {113.3217}& {7.5394}& {11.1784} \\
				\midrule
				$(\gamma,\delta,\eta)$ & $(50,20,-5)$ &  $(5,5,-2)$ & $(20,5,-2)$    \\
				\midrule
				BART &   {47.4422}& {5.4363} & {14.0281} \\
				Infinite BART &  {46.6487} &{4.6190} & {14.2574}\\
				\bottomrule
			\end{tabular*}
			\label{tab:mse.sim}
			\caption{Synthetic data from the model. Average MSE over 10 replicates. }
		\end{table}

		\paragraph{Real Data:} The out-of-sample prediction performance of BART and Infinite BART was evaluated on the following publicly available regression datasets:
		Airquality (153 observations, 6 variables), Boston Housing (506 observations, 14 variables), Iris (150 observations, 5 variables), Motor (626 observations, 7 variables), Mtcars (32 observations, 11 variables), Toxicity (546 observations, 9 variables). These datasets contain both covariate and categorical variables; for the latter, dummy variables were added for each level. Apart from removing missing values, no further pre-processing was applied. The MSEs over 10 train-test splits for each dataset are shown in Figure~\ref{tab:mse.real}, and Table~\ref{tab:mse.real} reports the corresponding average MSEs. In most cases, Infinite BART achieves predictive performance comparable to or better than classic BART. For the Airquality, the improvement is quite substantial with nearly a {45\%} decrease in MSE, and for Boston, Iris, {Motor} and Mtcars datasets, the reduction is approximately 20-30\% . 
		
		\begin{figure}
			\centering
			\includegraphics[height=7cm,width=1\linewidth]{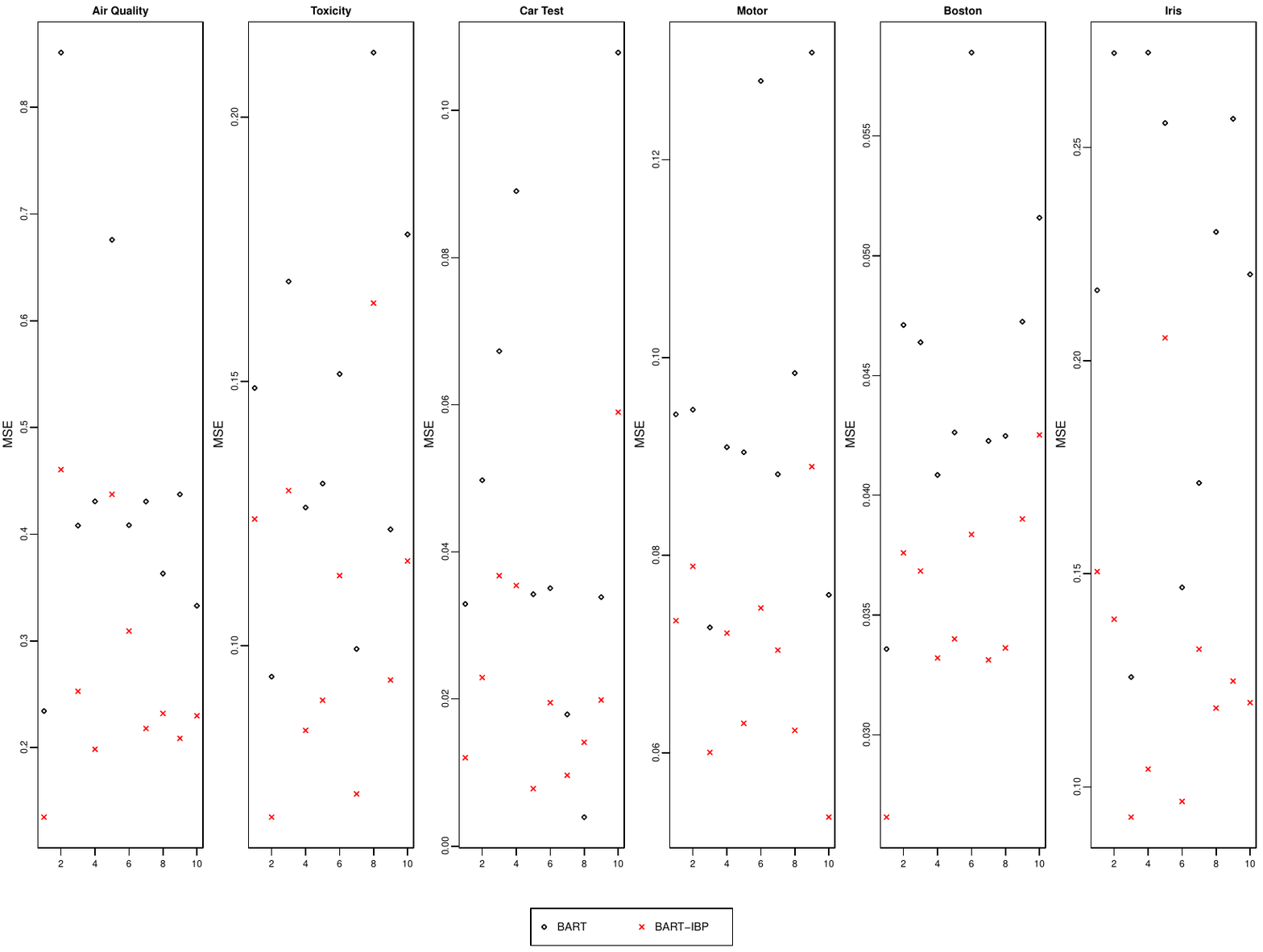}
			\caption{Real datasets. MSEs on 10 runs.  Black dots for BART; red crosses for Infinite BART.}
			\label{fig:mse.real}
		\end{figure}
		
		\begin{table}[ht]
			\centering
			\begin{tabular*}{28pc}{@{\hskip5pt}@{\extracolsep{\fill}}c@{}c@{}c@{}c@{}c@{}c@{}c@{\hskip5pt}}
				\toprule
				& {Air Quality}&Toxicity  &{Mtcars}   \\
				\midrule
				BART & {0.4573}& {0.0985} &{0.0472}\\
				Infinite BART & {0.2681}& {0.1054} &{0.0237} \\
				\midrule
				&  {Motor} & {Boston} & {Iris} \\
				\midrule
				BART  & {0.0964}& {0.0453}&{0.2167}\\
				Infinite BART & {0.0697}&  {0.0355}&{0.1284} \\
				\bottomrule
			\end{tabular*}
			\caption{Real Datasets. Average MSE over 10 replicates.}
			\label{tab:mse.real}
		\end{table}

		\paragraph{Clustered Data:} Finally, we consider simulated and real data, having some (soft) clustering in the regression function, with groups of responses having different regression functions. {Soft clustering is a form of clustering in which each observation has a vector of membership probabilities, rather than being assigned deterministically to one cluster. Thus an observation can effectively receive different contributions from several regression functions. For example, \cite{yang1997cluster} combine soft clustering with regression to account for heterogeneous regression relationships. Classical approaches for soft clustering include fuzzy $k$-means \citep{bezdek2013pattern,ferraro2024fuzzy}, finite mixture models and mixture-of-regressions, and mixture-of-experts models \citep{jordan1994hierarchical}, where different latent components are associated with different regression relationships. In particular, mixture-of-experts models allow the mixing probabilities to depend on covariates, providing a flexible framework for heterogeneous regression. Bayesian nonparametric extensions, including Dirichlet process mixture, allow the number of latent components to be inferred from the data. Therefore, infinite BART extends existing heterogeneous regression approaches by combining soft latent clustering with the flexible, nonlinear regression functions provided by BART, while allowing the number of effective regression components to adapt to the data.}
		
		A (soft) clustered extension of the Friedman example is as follows. We sample explanatory variables as $X_{i,j} \stackrel{ind}{\sim} \text{Beta}(\alpha_{j},\beta_{j})$, with parameters $\alpha_{j} = (\frac{j}{p+1})^{2}$ and
		$ \beta_{j} = \alpha_{j}(\frac{p+1}{j}-1)$, for $i=1,\ldots,n$ and $j=1,\ldots,p$. The mean and variance of variable $j$ are $E(X_{i,j})=\frac{j}{p+1}$ and $\text{Var}(X_{i,j})=1$. We assume $n=200$ and $K=5$ groups of observations, all equally sized $n/K=40$ . The responses of observations in the $k$-th group are sampled from
		\begin{equation*} 
			Y_{i}^{(k)} = 10\cdot\text{sin}(\pi x_{i,1 + (k-1)}x_{i,2+ (k-1)})+ 20(x_{i,3+ (k-1)} -0.5)^{2}+10x_{i,4+ (k-1)} +5x_{i,5+ (k-1)}+ \epsilon_{i}
		\end{equation*}
		for $i=1,\ldots, n/K$, and $k=1,\ldots,K$. Each of the $K$ groups has a Friedman regression function, but depending on different subsets of explanatory variables. Specifically, the first group uses variables $X_{1:5}$, the second group  variables $X_{2:6}$, the third group variables $X_{3:7}$, and so on. 
		
		Heterogeneity and (soft) clustering naturally occur in real datasets when one or more categorical or count variables that strongly influence the response are unobserved. Different levels of these variables induce distinct regression functions, effectively creating clusters of observations with varying responses. To illustrate this, we removed one significant factor or count variable in two real-data examples: `Species' variable was dropped from the Iris dataset, and `Constitutional indices' were removed from the Toxicity dataset.
		
		For both the reduced datasets and the Clustered Friedman example, BART and Infinite BART were run for 5,000 iterations, following a burn-in of 1,000 iterations, over 10 replicates. The average MSEs are reported in Table~\ref{tab:mse.clus}. In the Clustered Friedman example,  Infinite BART outperforms classic BART. For the two reduced real datasets, on the one hand, the average MSE of classic BART slightly deteriorates compared to that obtained using the full datasets (Table~\ref{tab:mse.real}), with an increase of 10-15\%, which can be explained by a reduction in predictability after dropping significant variables. On the other hand, the MSEs of Infinite BART remain nearly unchanged, highlighting its robustness to unobserved heterogeneity.

		\begin{table}[ht]
			\centering
			\begin{tabular*}{28pc}{@{\hskip5pt}@{\extracolsep{\fill}}c@{}c@{}c@{}c@{}c@{}c@{}c@{}c@{}c@{}c@{\hskip5pt}}
				\toprule
				& {Clust. Friedman} &  Toxicity  & {Iris}  \\
				\midrule
				BART  &  {33.0760} & {0.1438}&  {0.1895}\\
				Infinite BART &{28.6781}&{0.1072}&{0.1302}\\
				\bottomrule
			\end{tabular*}
			\caption{Real Datasets. Average MSE over 10 replicates. }
			\label{tab:mse.clus}
		\end{table}

		
		\subsection{Casual effects example}~\label{exp:casual}
		
		We consider a causal effect example, which includes nonlinearity and heterogeneous treatment effect in the regression function. The data is generated as follows:
		\begin{equation*}
			\begin{aligned}
				X_{i,1}&\sim \mathcal{N}\left(0,1\right), \ \ \ X_{i,2} \sim\text{Bernoulli}\left(0.5\right), \ \ \ X_{i,3},X_{i,4},X_{i,5} \sim U[0,1] \\
				T_{i} |\textbf{X}_{i}  &\sim \text{ Bernoulli}\left(\text{expit}\left(0.5+X_{i,1}-0.7X_{i,2}-0.3 \sin(2\pi X_{i,3})\right)\right)\\
				Y_{i}|T_{i},\mathbf{X}_{i} &\sim  T_{i} \times ( 1 + 0.5  X_{i,1}) + 2 + 0.3  X_{i,1}^2 - 0.5  X_{i,2} + \sin(2 \pi X_{i,3}) +\epsilon_{i} 
			\end{aligned}
		\end{equation*}
		where $T_{i}\in \{0,1\}$ is the treatment variable, and $\epsilon_{i}\stackrel{\text{iid}}{\sim} \text{N}(0,1)$. In this case, the true ATE is $E[ Y_{i}| T_{i}=1, \mathbf{X}_{i} ] - E[  Y_{i}| T_{i}=0, \mathbf{X}_{i}] = 1 +  0.5 \times E[X_{i,1}] = 1$. For each sample, we fit both Infinite BART and regular BART models for  5000 iterations, after 1000 iterations of burn-in, and without any thinning. The posterior predictive distribution of the ATE is obtained by comparing predictions of the outcome when setting  $T=1$ and $T=0$,  following the procedure described in subsection~\ref{sec:tasks}. As in the previous examples, we generated 10 independent replicates and computed the MSE of the ATE for each replicate, then averaged the MSEs across all runs. The average MSE is {0.35} for BART and {0.11} for Infinite BART, indicating that Infinite BART provides substantially more accurate treatment effect estimates in this causal setting. As demonstrated in \citet{H(11)},  BART is able to recover nonlinear relationships and higher-order interactions without requiring explicit model specification in the causal setting. Its structure enables it to adapt to complex response surfaces and capture heterogeneous treatment effects, while our proposed method further improves upon standard BART, yielding more accurate recovery of the underlying response surface.
		
		\section{Discussion} \label{sec:discussion}
		
		In this work, we proposed an extension of the classic BART model that allows the data to automatically determine the number of trees. The model further enables each observation to utilize only a subset of weak learners, capturing different features of the data and inducing (soft) clustering, where groups of observations can have distinct regression functions. The overall number of trees and their allocation across observations is governed by a weight matrix, which is inferred from the data via MCMC.  Most statistical tasks can be addressed using Infinite BART in a manner similar to the classic BART. Numerical experiments demonstrate the promising performance of the proposed method across variable selection, prediction, partial dependence estimation, and causal inference tasks.
		
		In terms of future research, the proposed model could be improved computationally and methodologically. From a computational perspective,  the MCMC sampler in Algorithm~\ref{algo1} has two main steps: 1) updating the trees and their parameters, and 2) updating the weight matrix $\mathbf{W}$. The computational cost of Step 1 per iteration depends on the number of active trees $K_{n}$, which, in our illustrative examples,  concentrates around values usually smaller than the default number of trees in BART. As a result, step 1 is generally less expensive in Infinite BART. However, Step 2 introduces additional computational cost, which does not scale well with the sample size $n$ (the number of rows in $\mathbf{W}$). An important computational improvement would be to implement the update of $\mathbf{W}$ more efficiently, for instance via sub-sampling or parallelization. Parallelization might be feasible by replacing the sequential representation of the IBP with its conditional Beta process representation \citep{TJ(07)}, and exploiting a stick-breaking representation to update $\mathbf{W}$ across rows simultaneously. Another potential improvement would be to develop a data-dependent initialization for the hyper-parameters $(\gamma,\delta,\eta)$. As discussed in subsection~\ref{sec:IBP} and Section~\ref{sec:applications}, the structure of $\mathbf{W}$ depends on these hyper-parameters. While a default choice produces dense matrices similar to classic BART, heterogeneous response surfaces may require longer burn-in periods if
		$\mathbf{W}$ is sparse. Therefore, data-informed initializations could possibly reduce convergence time and improve mixing.
		
		Methodological improvements are also possible. In Section~\ref{sec:applications}, Infinite BART performed well at in-sample tasks, particularly in variable selection. However, out-of-sample performance on real datasets showed improvements over classic BART only in some cases, and was comparable in others. One possible explanation is that the prior on matrix $\mathbf{W}$ does not directly depend on the explanatory variables, and these only influence the posterior updates of $W_{i,\cdot}$ in Step 3 of Section~\ref{sec:mcmc}. Incorporating a covariate-dependent prior for $\mathbf{W}$ could improve predictive performance and capture structure more effectively. Finally, while Infinite BART was proposed as an extension of the classic BART model, the idea of introducing a weight matrix to allow the data to select the number of trees  could potentially be applied to other generalizations of BART.

\bibliographystyle{chicago}

\bibliography{../references}

\end{document}